 \documentclass{emulateapj}

\newcommand{\xte}{{\it RXTE}}

\newcommand{\epcs}{{\rm ergs\,cm^{-2}\,s^{-1}}}

\newcommand{\src}{4U~1728$-$34}

\slugcomment{Accepted by ApJ}

\shorttitle{Radius-expansion burst spectra from 4U~1728$-$34}
\shortauthors{Galloway et al.}

\begin{document}

\title{Radius-expansion burst spectra from 4U 1728$-$34: an ultracompact
binary?}

\author{Duncan K. Galloway\altaffilmark{1},
   Yangsen Yao\altaffilmark{2},
   Herman Marshall\altaffilmark{3},
   Zdenka Misanovic\altaffilmark{1},
   Nevin Weinberg\altaffilmark{4}}

\email{Duncan.Galloway@monash.edu}

\altaffiltext{1}{Center for Stellar and Planetary Astrophysics and School
of Physics,
  Monash University, VIC 3800, Australia}
\altaffiltext{2}{Center for Astrophysics and Space Astronomy, University
of Colorado, 389 UCB, Boulder, CO 80309}
\altaffiltext{3}{Kavli Institute for Astrophysics and Space Research,
Cambridge MA 02139}
\altaffiltext{4}{University of California, Berkeley, CA 94720}

\begin{abstract}
Recent theoretical and observational studies have shown that ashes from
thermonuclear burning may be ejected during radius-expansion bursts,
giving rise to
photoionisation edges in the X-ray spectra.  We report a search for such
features in {\it Chandra}\/ spectra 
observed from the low-mass X-ray binary 4U~1728-34.
We analysed the spectra from four 
radius-expansion
bursts detected in 2006 July, and two in
2002 March, but found no evidence for discrete features. We estimate upper
limits for the equivalent widths of edges of a few hundred eV, which for the
moderate temperatures observed during the bursts, are comparable with the
predictions.
During the 2006 July observation 4U~1728$-$34
exhibited weak, unusually frequent bursts (separated by $<2$~hr in some
cases), 
with
profiles and $\alpha$-values characteristic of hydrogen-poor fuel.
Recurrence times as short as those measured are
insufficient to exhaust 
the accreted hydrogen 
at solar composition, 
suggesting that the source accretes
hydrogen deficient fuel, for example from an evolved donor.
The detection for the first time of a 10.77~min periodic signal in the
persistent intensity, perhaps arising from orbital modulation, supports
this explanation, and suggests that this system is an ultracompact binary
similar to 
4U~1820$-$30.
\end{abstract}


\section{Introduction}

One of the highest priorities for observational studies of neutron stars
is measurement of the mass and radius, sufficient to constrain the
uncertain equation of state \cite[EOS; e.g. ][]{lp07}. A
promising avenue to achieve such measurements is to detect surface
spectral features from accreting neutron stars, which may preferentially
show such features during thermonuclear (type-I) bursts, caused by
unstable ignition of accreted H/He on the surface of the neutron star.
Identifying such features allows measurement of the surface redshift, and
hence the compactness ($M/R$ ratio), which in turn allows constraints to
be placed on the EOS. In recent years,
only one claim for such features has been made \cite[]{cott02}, and
subsequent efforts have not been able to confirm this result
\cite[e.g.][]{cott08}\footnote{Additionally, the case for slow rotation in
this system (required to produce lines as narrow as those observed) has
weakened, with the detection of burst oscillations at 552~Hz
\cite[]{gal10a}.}. The
majority of observational efforts to date have focussed on sources which show
frequent bursts, which tend to be faint.
Another class of bursts are significantly brighter, since the flux
during the burst rise exceeds the 
Eddington limit \cite[e.g][]{lew93}, at which
point the radiation pressure of the burst luminosity exceeds the local
gravity.
Recent theoretical work suggests that
such bursts could drive a wind containing heavy-element ashes from the
burning, which would 
imprint absorption features on the X-ray spectrum at the peak of the burst
\cite[]{wbs06}. 
Observational support for this hypothesis has been found from the
low-resolution spectra of extremely intense ``superexpansion'' bursts.
These bursts are the most intense of those 
thought to be powered by H/He, in contrast to the even longer, infrequent
``superbursts'', which are thought to be powered instead by carbon
\cite[e.g.][]{cumming06}. In superexpansion bursts observed by the
{\it Rossi X-ray Timing Explorer} ({\it RXTE}), 
\cite{zand10a} found highly significant residuals for fits with a blackbody
model. The residuals could be explained by absorption edge features, with
energies consistent with highly-ionised iron-peak elements and depths
indicative of abundances $\gtrsim 100$ times solar.  However, the
relatively poor spectral resolution 
prevents identification of the features.

Here we describe an attempt to detect and measure the X-ray spectrum at
the peak of 
radius-expansion bursts using the {\it Chandra}\/ High
energy Transmission Grating Spectrometer (HETGS).
Few sources reliably show frequent, radius expansion bursts, and the key
challenges for this search are the identification of a target and the
observation triggering. 
One candidate is the globular cluster source 
4U~1820$-$30 which orbits in
an ultracompact binary \cite[$P_{\rm orb}=685$~s][]{swp87} and thus likely
accretes almost pure helium.  This system only exhibits bursts when in the
minimum of its $\approx180$-d intensity cycle \cite[e.g.][]{pt84}, and on
one occasion was observed to produce 7 radius-expansion bursts separated
by 193~min on average \cite[]{haberl87}.  However, frequent bursts are not
consistently observed, and recent pointed observations targeting the
long-term minima have had limited success, for example detecting only four
bursts in 1~Ms of {\it Rossi X-ray Timing Explorer}\/ observations
\cite[e.g.][]{bcatalog}.
A better target is 4U~1728$-$34, a well-known burst source for which the
accretion rate 
does not reach such extreme values, and is 
typically $\sim0.1\dot{M}_{\rm
Edd}$.  This rate is substantially lower 
(by almost an order of magnitude)
than the maximum reached by
4U~1820$-$30 and 4U~1728$-$34 commonly
produces radius-expansion bursts every $\approx4$--5~hr. The
burst properties are otherwise similar to those of 
4U~1820$-$30
\cite[]{cumming03}, and as we show here, it is possible that
4U~1728$-$34 is also an ultracompact system.

\begin{deluxetable*}{cllllccl}[t]
\tablecaption{X-ray observations of \src\label{obslog}}
\tablewidth{0pt}
\tablehead{
  & & & &
\colhead{Data} & \colhead{Exposure} & \colhead{No. of} \\
\colhead{No.} & \colhead{Date} & \colhead{Instrument} & \colhead{obsid} &
\colhead{mode} & \colhead{(ks)} & \colhead{bursts} & \colhead{Ref.} \\
}
\startdata
1 & 2002 Mar 4 & {\it Chandra} & 2748 & TE & 29.4 & 2 & [1] \\
2 & 2002 Oct 29 & {\it XMM-Newton} & 0149810101 & Timing (PN) & 28.1 & 0 \\
3 & 2006 Jul 17--18 & {\it Chandra} & 6568 & CC & 49.5 & 4 & [2, 3] \\
4 & 2006 Jul 18--20 & {\it Chandra} & 6567 & CC & 151.8 & 18 & [2, 3] \\
5 & 2006 Jul 22-23 & {\it Chandra} & 7371 & CC & 39.7 & 3 & [2, 3] \\
\enddata

\tablerefs{1. \cite{dai06}; 2. this paper; 3. 
\cite{misanovic10} }

\end{deluxetable*}

\section{Observations \& analysis}

We observed \src\ between 2006 July 17--23 with the 
HETGS \cite[]{hetgs05} aboard {\it
Chandra}. The HETGS consists of two
separate grating arrays (the Medium- and High-Energy Gratings, MEG and
HEG) which intercept about 40\% of the incoming photons and disperse them
linearly along the Advanced CCD Imaging Spectrometer (ACIS). These
instruments together provide spectra in the range 1.2-31~\AA\ (10-0.4 keV)
with spectral resolution ($E/\Delta E$) of up to 1000, and wavelength
resolution of up to 0.012~\AA. We adopted the continuous clocking (CC)
observing mode, for which the ACIS-S CCDs are read out continuously. This
sacrifices one axis of  spatial imaging but allows much higher incident
photon countrates before photon pileup (multiple photons arriving at the
same pixel in the same frame time) becomes a problem. Such mitigation
strategies are necessary to deal with the highly variable incident count
rates for a bursting source, but complicates the data reduction and
spectral analysis. Specifically, CC-mode observations prevent the usual
order sorting for dispersed spectra, which rely (in part) on both spatial
axes, although the photon energy measured by the CCD allows unambiguous
order sorting, at least for low dispersion orders. CC-mode observations
also prevent extraction of the usual background spectrum.

We reduced the {\it Chandra}\/ data with {\sc ciao} version 4.0 \cite[December
2007;][]{ciao06}, using the calibration database ({\sc CALDB}) version
3.4.2. We undertook spectral fitting with {\sc isis} version 1.4.9-4
\cite[]{isis00}. We calculated response matrices (using {\sc mkgrmf}) and
ancilliary response files (using {\sc fullgarf}) for each observation.
We also used the response and ancilliary files calculated for the entire
observation to analyse the spectra from any bursts detected in those
observations.

We also analysed archival {\it Chandra}\/ timed exposure (TE) mode
observations of 4U~1728$-$34 made on 2002 March 4, as well as archival {\it
XMM-Newton}\/ observations from 2002 October 29. We analysed the {\it
XMM-Newton}\/ data using the Science Analysis Software ({\sc SAS}) version
7.0.0 (June 2006). We used data from the European Photon Imaging Camera
(EPIC) pn CCD camera \cite[]{kuster02}, which was operated in timing mode.
The EPIC pn camera is sensitive to photons in the range 0.15--15~keV, and
with an effective area peaking at around 1400~cm$^2$ at 2~keV. As with
CC-mode for {\it Chandra}, in pn timing mode the central CCD alone is read
out continuously, providing an effective time resolution of 0.03~ms.
The observations used in this paper are summarised in Table \ref{obslog}.

\section{Results}
\label{sec1}

We obtained in 2006 July three separate pointings of approximately 50, 150
and 40~ks each, spanning almost 6~days, and detecting four, eighteen, and
three bursts in each segment (respectively). We first analysed the
persistent
source spectrum between the bursts, 
to establish the persistent X-ray intensity.
The raw lightcurve indicated significant variations in the persistent
count rate from observation to observation, and also between each
inter-burst interval.
Thus, we divided each observation into
segments delineated by each pair of bursts, excluding the intervals up to
150~s after each burst and 50~s before the next burst. For the first
bursts of each pointing, we integrated from the beginning of the pointing
to 50~s before the burst. We then fit the summed, first-order MEG and HEG
spectra simultaneously in the
energy range 
2.1--8.3~\AA
(6--1.5~keV) with
an absorbed comptonisation model, motivated by earlier broad-band
spectroscopic measurements
(see 
\citealt{misanovic10}, hereafter M10, for full details of
the persistent spectral fitting, and references theirein for 
additional analyses of the persistent spectrum seen by {\it Chandra}).
The
neutral column density was fixed at 
$2.29\times10^{22}\ {\rm cm^{-2}}$.

Like many low-mass X-ray binaries (LMXBs), \src\ shows substantial
variation in its X-ray spectrum and intensity on a range of timescales,
from days to decades \cite[e.g.][]{lew93}, and the spectral ``states''
influence the burst properties. 
The integrated persistent flux during the 2006 {\it Chandra}\/
observation was in the range 
1.16--$2.30\times10^{-9}\ \epcs$ (1.5--6~keV),
equivalent to at least 3--$5\times10^{-9}\ \epcs$ in the 2.5--25~keV band,
based on the extrapolated spectral model.
These values are atypically high for the source; extensive
earlier observations with 
{\it RXTE} 
found the source in this flux range only 10\% of the time
\cite[]{bcatalog}. 
In the absence of spectral information above 10~keV, it is difficult
to identify the source spectral state; however, the nearest \xte\/
observation, on 2006 July 24\footnote{Observation ID 92023-03-72-00},
found a characteristically ``soft'' spectrum (e.g. with electron
temperature for a {\tt compTT} spectral component of 4.6~keV), and X-ray
colors indicating the source was in the lower ``island'' region of the
X-ray color-color diagram \cite[see e.g.][]{bcatalog}. 
The source
flux was highest during the longest (second) pointing, consistently above
$2.0\times10^{-9}\ \epcs$ (1.5--6~keV). 
In contrast to previous observations \cite[e.g.][]{dai06}, no iron lines were detected in the
persistent spectrum. For a distance to the source of 5.2~kpc \cite[from the
peak flux of radius-expansion bursts observed by {\it RXTE};][]{bcatalog},
and a bolometric correction 
which assumes a characteristic high-state ``soft'' spectrum \cite[with electron
temperature of 3~keV, c.f. with][]{falanga06},
the
corresponding range of accretion rate is 
4--7\%\footnote{Note that the fraction will be correspondingly higher
for nonzero H-fraction, for example
6--11\% of the Eddington rate
$\dot{M}_{\rm Edd,H}=10^{18}\ {\rm g\,s^{-1}}$, 
assuming H-fraction in the accreted fuel of $X_0=0.7$.} of the Eddington rate
($\approx1.8\times10^{18}\ {\rm g\,s^{-1}}$, assuming pure He accretion).

\subsection{The X-ray bursts}
\label{bursts}

\begin{figure*}
 \epsscale{1.15}
 \plotone{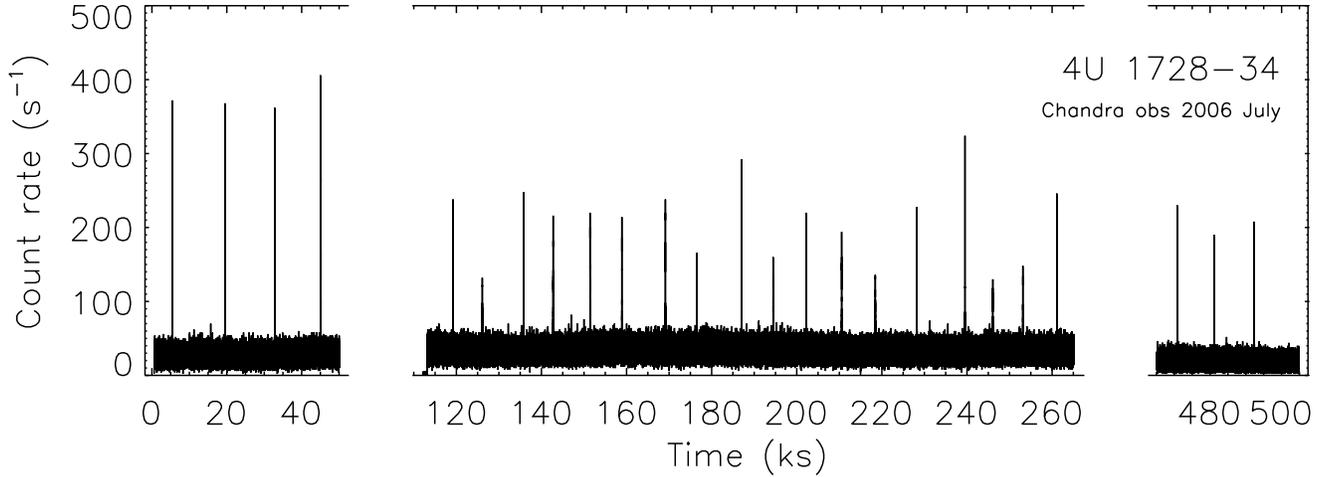}
 \figcaption[]{The 2006 July {\it Chandra}\/ observation of \src, showing
the 25 thermonuclear bursts detected, which were separated by recurrence
times in the range 1.82--3.92 hr. The first-order photons dispersed by the
HETGS were combined and binned at 0.5~s time resolution; the lightcurve
for the zeroth-order (undispersed) photons was dominated by pileup, which
resulted in truncation of the peaks of the bursts, so these photons were
excluded.
 \label{fbursts} }
\end{figure*}

\begin{figure}
 \epsscale{1.15}
 \plotone{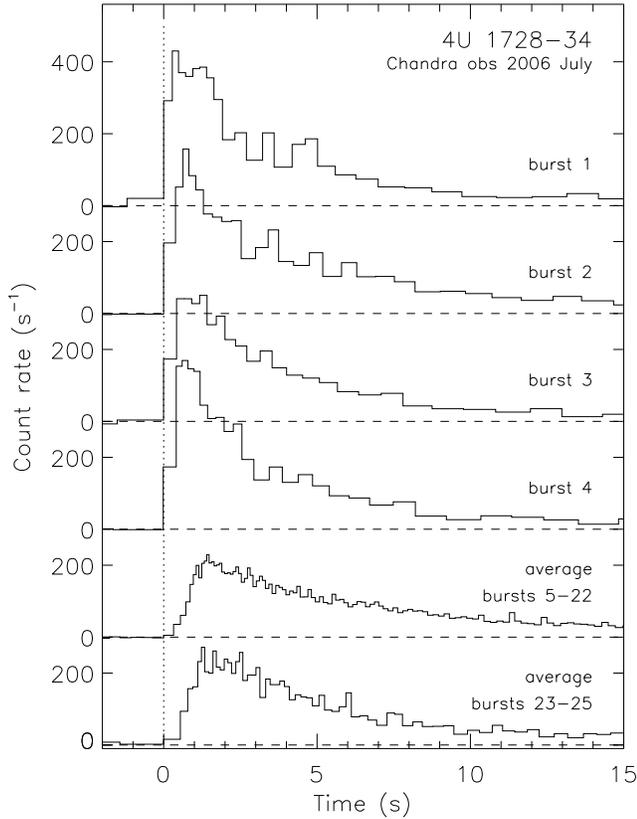}
 \figcaption[]{Burst profiles from the three intervals comprising
the 2006 July {\it Chandra}\/ observation of \src. The first four bursts,
from observation 6568, are plotted individually; we subtract the pre-burst
emission ({\it dashed lines}) and shift the profiles vertically for
clarity. Dispersed events were binned to ensure 100~counts per bin. We
also plot the combined lightcurves from second and third observations,
6567 and 7371, binned with 400 and 100~counts per bin, respectively. 
Note the markedly different peak intensity and shape of bursts 1--4
compared to those that followed.
 \label{meanprofiles} }
\end{figure}

We detected 25 bursts during the 2006 July observations
(Fig. \ref{fbursts}).
The four bursts in the first observation
(\#6568) were separated by $3.7\pm0.3$~hr on average, while in the second
and third observations (\#6567, 7371) the mean recurrence time 
was significantly shorter, at $2.4\pm0.5$~hr.
The bursts themselves were short, with rise times of $\approx1$~s and
lasting only $\approx15$~s (Fig. \ref{meanprofiles}). The four bursts in
the first segment (observation \#6568) were the brightest, with the bursts
in the second and third segments reaching count rates only approximately
2/3 the peak as in the first segment.
A more detailed discussion of the burst properties
and the inferred ignition conditions can be found in M10.

\begin{figure}
 \epsscale{1.15}
 \plotone{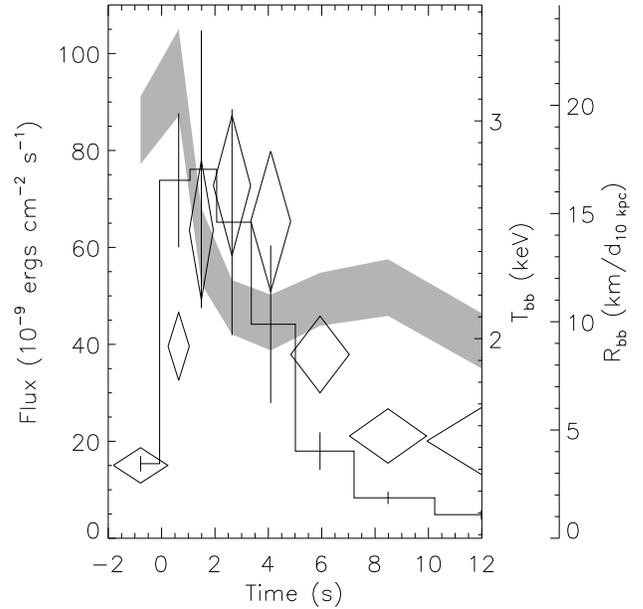}
 \figcaption[]{Time-resolved spectroscopy of the summed signal for the
first four bursts, detected in observation \#6568.
The histogram shows the (bolometric) burst flux (left-hand
$y$-axis) in units of $10^{-9}\ {\rm erg\,cm^{-2}\,s^{-1}}$, with error
bars indicating the $1\sigma$ uncertainties. The grey
ribbon shows the $1\sigma$ limits of the blackbody radius (outer
right-hand $y$-axis) in ${\rm km}/d_{\rm 10kpc}$. The diamonds show
the $1\sigma$ error region for the blackbody temperature (inner right-hand
$y$-axis) in keV. 
Note the elevated blackbody radius (to approximately twice the value in
the tail) during the second time-bin near the burst peak. This radius
increase 
indicates 
photospheric radius-expansion in these bursts.
 \label{timespec} }
\end{figure}

Two of the bursts in the first observation (\#6568) had indications of a
constant intensity at maximum, lasting $\approx1$~s. Since constant
luminosity in the peak is an indication of radius-expansion, 
we carried out time-resolved blackbody spectral fits of the combined 
first-order 
HEG and MEG 
data
for each set of bursts in each observation, to test for the characteristic
variations in the blackbody radius and temperature expected for
radius-expansion bursts. Following the 
conventional approach, we
subtracted the pre-burst emission as background, and estimated the
bolometric flux based on the blackbody spectral parameters.
The average peak flux for the bursts in the three observations (\#6568,
6567 and 7371) was
$8\pm2$, $7\pm2$ and $(5.9\pm1.9)\times10^{-8}\ \epcs$, respectively.
We found evidence for an elevated blackbody radius, coincident with a
decrease in the blackbody temperature, only for the first four bursts
(Fig. \ref{timespec}). While we did not fully resolve the rise in
blackbody radius, this may largely have been due to the relatively low
count rate early in the burst coupled with the steep rise time. 
These variations 
suggest that the first
four bursts exhibited radius-expansion.
Although the confidence interval for the mean fluxes for the subsequent
bursts were consistent with the maximum reached by
radius-expansion bursts observed by other instruments \cite[]{gal03b}, the
time-resolved spectral variations in those bursts 
gave no indications of radius-expansion.

We measured the ratio $\alpha$ \cite[e.g.][]{gottwald86} of the integrated persistent
flux within each interval between the bursts, to the integrated flux of
the subsequent burst (from time-resolved spectral fits; see M10)
and obtained values
which were 
190
in the mean.
The rapid timescales of the bursts, as well
as the high $\alpha$-values, are both consistent with ignition of primarily He
fuel. Although both the persistent flux and the burst recurrence times
varied significantly, we found no correlation between the two, as might be
expected (under the usual assumption that the persistent flux tracks the
accretion rate).
As discussed by M10, this lack of correlation may result from incomplete
burning, or alternatively a varying area over which accretion takes place
on the neutron star.

\subsection{Search for spectral features during the bursts}
\label{burstspec}

We made a search for discrete features in the spectra from the
bursts observed by {\it Chandra}\/ in 2006 July.
We first focussed on the four bursts observed in the first observation,
\#6568, since only these exhibited any evidence for photospheric
radius-expansion.
We extracted spectra from various time intervals of each burst, and
combined the $\pm1$ orders to obtain a single first-order spectrum for
each grating arm (HEG and MEG). We searched spectra extracted from 0--2~s
(relative to the burst start time), and 0--4~s.
The combined total counts in HEG/MEG first-order for each of
these intervals was 
926/912 and	
1740/1591.	
We grouped the bins by various factors to obtain bin widths of 0.01, 0.03,
0.05, 0.11 and 0.22~\AA, in order to search for both broad and narrow
features. Since the counts per bin for these spectra was 
$<10$ at short and long wavelengths (away from the peak sensitivity) we
adopted Gehrels weighting to determine errors on each bin,
which better approximates Poisson uncertainties for low counts. This is
preferable over the alternative approach, i.e. to bin adaptively to ensure
some minimum number of counts per bin, since absorption features that
saturate the spectrum and reduce the counts to zero will be missed.
We fit each pair (HEG and MEG) of first-order spectra using an absorbed
blackbody with column density $n_H$ fixed at $2.3\times10^{22}\ {\rm
cm^{-2}}$ (derived
from fits covering the first 2 seconds of each burst; see also M10). We
set a detection threshold of 3-$\sigma$, taking into account the total
number of bins (HEG+MEG) in each time interval and binning.

\begin{figure}
 \epsscale{1.13}
 \plotone{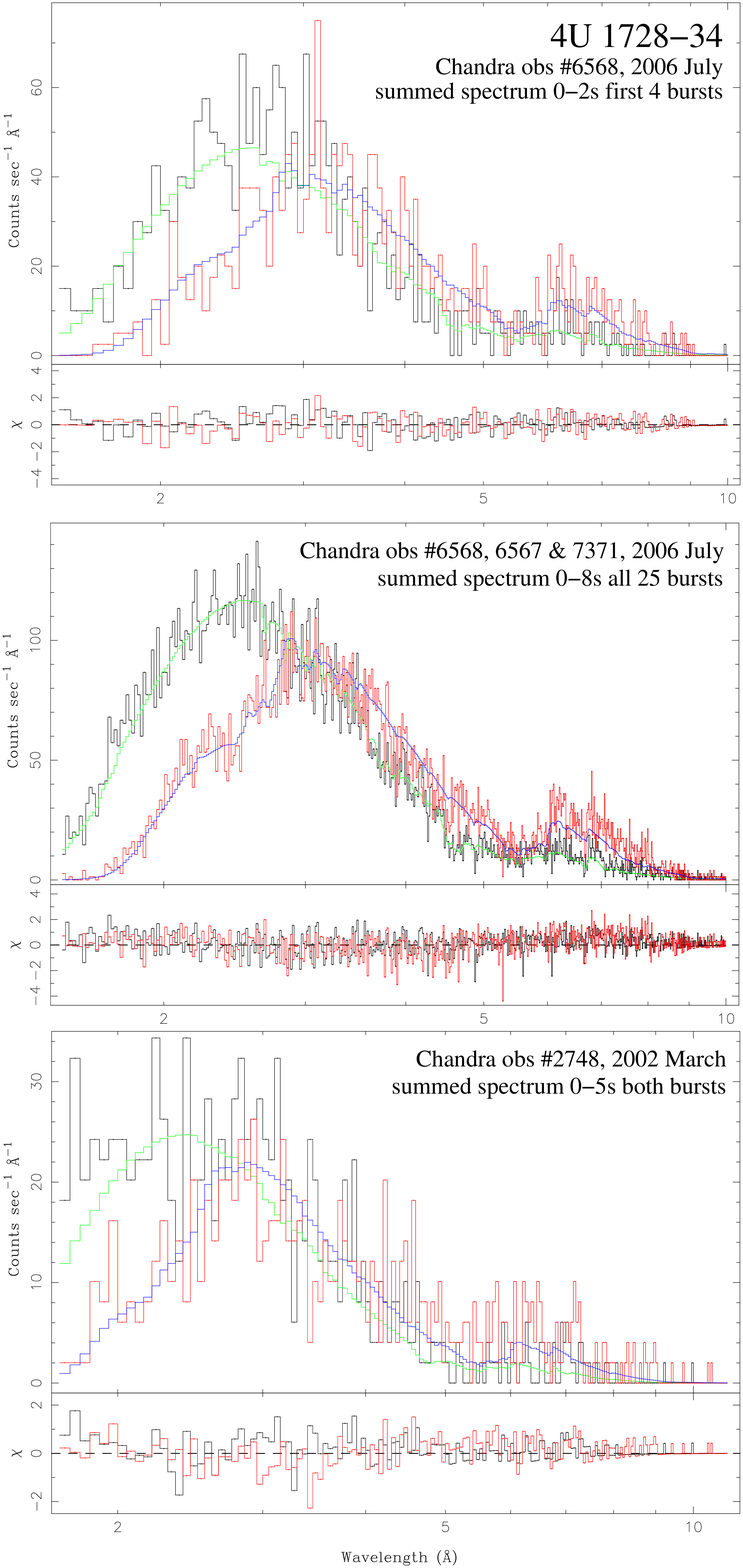}
 \figcaption[]{HEG and MEG first-order spectra extracted from
thermonuclear bursts observed by {\it Chandra}. The top panel shows the
spectra extracted from 0--2~s (relative to the burst start) from the first
four bursts of observation \#6568, in 2006 July, and binned at 0.11~\AA.
The summed $\pm1$ order HEG and MEG are shown separately ({\it black and
red histograms}), as well as the best-fitting blackbody model ({\it green
and blue}). 
The middle panel shows the spectra extracted from 0--8~s of all 25 bursts
observed by {\it Chandra}\/ in 2006 July, binned at 0.03~\AA. Note the
feature around 5.28~\AA, which is closest to our $3\sigma$ significance
threshold, taking into account the number of spectral bins.
The bottom panel shows the spectra extracted from 0--5~s of the two bursts
observed by {\it Chandra}\/ in 2002 March (observation \#2748). 
 \label{spectra} }
\end{figure}

We found no significant deviations from the blackbody (continuum) for any
choice of binning, in either time interval. For the spectra extracted from the
first two seconds of the bursts, which likely covers the entire radius
expansion interval, we derive $3\sigma$ upper limits for
emission/absorption lines of equivalent width in the range 100--180~eV. We
show the spectrum extracted over this interval in Figure \ref{spectra},
top panel.

We also estimated the upper limits on detection of absorption edges in the
spectra. We added edge components to the model with the edge energy fixed
at 
$\approx3.1$~\AA (4~keV), roughly where the observed count rate per bin of both HEG and
MEG reached a maximum, in order to measure the best achievable limit with
the available data. We
found a $3\sigma$ upper limit of 0.26 (0.20) on the optical depth $\tau$ for
the 0--2~s (0--4~s) spectra respectively, corresponding to equivalent
widths of 280~(250)~eV.
We also searched spectra from after the
radius expansion episode, in case the photosphere was too hot during 
radius-expansion for neutral or incompletely-ionised species to survive. In spectra extracted from
time window covering 2--16 and 4--16~s after the burst start, we obtained
upper limits as before on edges of 250 and 260~eV, respectively.

Although the bursts observed in the second and third observations (\#6567
and 7371) likely did not exhibit radius-expansion, we undertook a second
search of spectra extracted from a wider choice of time intervals
(relative to the start) of all 25 bursts.
The combined total counts in HEG/MEG first-order for each of
these intervals, i.e. 0--2, 0--4, 0--8 and 0--16~s, as well as 
2--16, 4--16 and 8--16~s was 
3540/3350,	
7670/7180,	
12880/12470,	
18680/19190,	
respectively.
We found no broad or narrow features that exceeded our $3\sigma$
confidence threshold for any choice of time interval or binning. The
most significant deviation was in the MEG spectra extracted from 0--8~s of
all 25 bursts, and binned at 0.03~\AA, in which a single bin around
5.28~\AA\ was below the model fit at a significance of almost $3\sigma$
(Fig. \ref{spectra}, middle panel). No similar feture was seen in the HEG
spectrum.

Despite several attempts in recent years, observations of radius-expansion
bursts by {\it Chandra}\/ or {\it XMM-Newton}\/ have eluded observers. In
fact, we are aware of only one other confirmed radius-expansion burst, which was
detected on 2002 March 4
22:42:45
(MJD~52337.94637) in observation \#2748 (Table \ref{obslog}). This burst
was also detected by {\it RXTE}\/ \cite[burst \#101 in][]{bcatalog},
and was found to exhibit photospheric radius expansion. This event was
preceded by a similar burst at 
UT 18:35:24 (MJD~52337.77459), 4.12~hr earlier. Under the assumption that
this burst too reached the Eddington limit, we extracted spectra from the
peak and the tail of each burst, and combined them to search for discrete
spectral features.

Unlike the observations in 2006 July, observation \#2748 was performed in
Timed Exposure (TE) mode, with the CCDs read out every 1.44~s. The extreme
brightness of the source during the radius expansion bursts likely led to
substantial pileup during the bursts, although this was not obvious
from the dispersed burst spectra. We combined spectra from the HEG and MEG $\pm1$
order from the first five and next 14--19 seconds of the two bursts, and
examined the spectra for any evidence of discrete features. Due to the
small number of total counts in the two sets of spectra (600--800 counts
total in each grating arm), we rebinned the HEG (MEG) by a factor of 20
(10) to achieve a bin size of approximately $0.11$~\AA. We obtained an
acceptable fit using an absorbed blackbody model, and found no discrete
features which exceeded our $3\sigma$ detection threshold (taking into
account the number of spectral bins). A representative spectrum, from the
first five seconds of the two bursts summed, is shown
in Fig. \ref{spectra} (bottom panel).
We estimate $3\sigma$ upper limits on the equivalent width of discrete
features of $\pm150$~eV.

\subsection{Persistent intensity variations}
\label{orbit}

The long, uninterrupted X-ray lightcurve accumulated during the {\it
Chandra}\/
observation allowed a search for periodic signals 
with sensitivity greater than for any other dataset, particularly at
low energies.
Persistent X-ray intensity variations in LMXBs are sometimes
observed at the orbital period, which for \src\ remains unknown. We
concentrated on the longest (second) segment of data, and created a
lightcurve from first-order photons (both HEG and MEG) in the
$\approx6$--25~\AA (2--0.5~keV) energy range (where orbital variations of neutral
absorption would be expected to have the largest effect) at 60~s time
resolution. We removed the 1--2~min segments including each burst, and
filled the resulting gaps with normally-distributed values consistent with
the local mean and standard deviation. The overall mean countrate for the
lightcurve was $(8.0\pm0.4)$~count~s$^{-1}$.

\begin{figure}
 \epsscale{1.2}
 \plotone{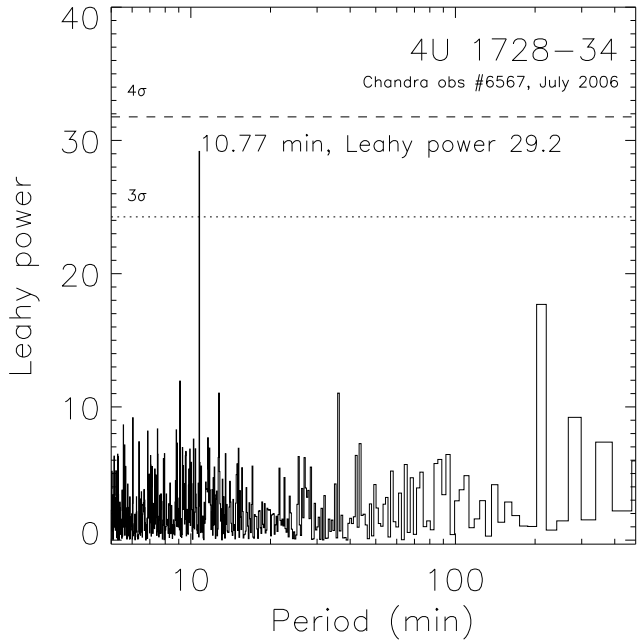}
 \figcaption[]{Power density spectrum for the longest contiguous segment
of the 2006 {\it Chandra}\/ observation of \src. The 3- and 4-$\sigma$ detection
thresholds  (taking into account the number of bins in the frequency range
of interest) are indicated, as is the bin with the maximum power, at a
period of 10.77~min. 
 \label{fft} }
\end{figure}

The Fourier power spectrum of the resulting light curve exhibited a peak
with Leahy power of 25--29
at 10.77~min, significant (taking into account the number of trials) at
between 3 and $4\sigma$ (i.e. at greater than 99.7\% significance; Fig.
\ref{fft}). The peak power and significance vary because of the use of
random values
to fill the gaps in the lightcurve where the bursts were removed, but the
detection was
consistently $>3\sigma$ over many trials.  The signal was approximately
sinusoidal, with an RMS amplitude of 0.5\%.
We searched for the 10.77~min signal in the full-range lightcurve, as
well as other energy bands. The Leahy power at the corresponding frequency
in the FFTs from the full-range lightcurve was in the range 17--20,
indicating a $\approx2\sigma$ detection. There was evidence that the
signal power decreased with energy above 2~keV; nominal power was detected
both in the 2--3 and 3--4~keV bands, with corresponding 3-$\sigma$ upper
limits on the RMS amplitude of 0.4 and 0.3\%, respectively. A stronger
signal with Leahy power $\approx40$ was detected in the 3--4~keV
lightcurve at a period of 16.64~min, very close to the spacecraft yaw
dither period of 1000~s, and so likely instrumental.
We also searched a lightcurve comprising the zeroth-order photons in the
$<2$~keV band. The count rate in the undispersed low-energy photons was
just 2.2~count~s$^{-1}$ in the mean, substantially less than that of the
first-order photons, likely due to pileup. We detected no significant
power at the 10.77~min period, deriving an upper limit on the rms
amplitude of 0.7\%, which cannot rule out the detection in the dispersed
photon lightcurve.

Several other archival observations of \src\ are available, and we also
undertook timing analyses of some of these data in an attempt to confirm the
10.77~min period. 
The earlier {\it Chandra}\/ observation on 2002 March 4 \cite[observation
ID \#2748, Table \ref{obslog}; see also ][]{dai06} found the source at a
significantly lower intensity, and the first-order count rate below 2~keV
was around
5.3~count~s$^{-1}$. We computed a 60-s lightcurve omitting the intervals
surrounding the two bursts (see \S\ref{burstspec}), and filled the
resulting gaps with random values as with the 2006 data.  No
signal was detected at the candidate period. With the observation only
lasting 9.1~hr, the corresponding upper limit on the candidate signal
amplitude was 1.7\% ($3\sigma$).
An {\it XMM-Newton}\/ observation on 2002 October 29 (observation ID
\#0149810101, Table \ref{obslog}) was similarly substantially shorter than
the 2006 {\it Chandra}\/ observation, at only 7.5~hr. We calculated a
lightcurve by accumulating photons with energies $\le 2$~keV in 60-s bins,
with a mean rate of 9.4~count~s$^{-1}$. We found no signal at the
candidate period, and derive a $3\sigma$ upper limit on the rms amplitude
of 0.8\%. As with the earlier {\it Chandra}\/ data, this is above the
detected amplitude of the signal in the 2006 observations.

\subsection{The X-ray position and optical counterpart}

The detection of a short-period signal in the persistent X-ray emission
during the 2006 observation led us to reexamine the optical
counterpart proposed by \cite{marti98}. The infrared magnitudes combined
with the measured column density lead those authors to suggest a
counterpart ``less luminous and less massive than a middle B/early A V
star or a F-K III star'', taking into account the substantial contribution
of the accretion disk. 
Here we investigate whether
the observations are consistent with an ultracompact counterpart similar
to that of the similar 10-min binary 
4U~1820$-$30.

The counterpart to that system was detected in the $B$ and F140W UV bands by
\cite{king93}. The ratio of the absorption-corrected fluxes in the two
bands was approximately $(\Delta\lambda/\lambda)^{-3.2}$, so that the
spectral slope was close to the Rayleigh-Jeans value. 
We extrapolated the
observed spectrum to the $J$ and $K$ bands, and corrected for the different
distances of 7.6~kpc for 
4U~1820$-$30 and 5.7~kpc for 4U~1728$-$34
\cite[assuming that radius-expansion bursts reach the Eddington limit for
a helium atmosphere;][]{gal03b}. The expected magnitudes at the distance of
\src\ are $J=17.5$ and $K=17.3$, which are substantially different to the
absorption-corrected values of 19.6 and 15.1 measured by \cite{marti98}.
On the other hand, if we extrapolate using a spectral slope of 4, we
expect $J=19.4$ and $K=19.7$, which at least matches the $J$-band
luminosity for the proposed counterpart. We conclude that a
counterpart similar to 
4U~1820$-$30 may give a consistent $J$-band
luminosity to the proposed \src\ counterpart, but the measured $K$-band
brightness is too high. This may indicate an additional IR component as
has been inferred to exist in other compact binaries \cite[e.g.][]{gre06}.

Alternatively, we consider the possibility that the proposed counterpart
is unrelated and positionally coincident with \src\ by chance. First we
determined an accurate X-ray position from the {\it Chandra}\/ observation
\#2748 (Table \ref{obslog}). The zeroth-order image for this bright source
was significantly piled up, leaving a ring of emission rather than the
point source usual for fainter sources. The standard source detection
algorithms cannot provide a precise position for such images, so instead
we fit the position of the dispersed HEG and MEG spectra, as well as
the readout trace (an artifact arising from the shuffling of charge along
the CCD illuminated by a bright point source) and computed their
intersection. We first checked for, and corrected, an aspect offset of
$-0\farcs46$ in R.A. and $-1\farcs03$ in Dec\footnote{This correction was
required since we used data processed prior to 2004; see {\url
http://cxc.harvard.edu/cal/ASPECT/fix\_offset}}. We determined a position
for \src\ of 
$R.A. = 17^{\mathrm h}31^{\mathrm m}57\fs69$,
decl. = $-33\arcdeg50\arcmin01\fs3$ (J2000.0),
with an estimated 90\% confidence uncertainty of $0\farcs6$ (corresponding
to the systematic pointing uncertainty for {\it Chandra}). We note that
this position is within $1\farcs3$ of the radio source \cite[]{marti98},
confirming the positional coincidence of the radio and X-ray sources.

Finally, we estimated the field density of stars with $K\leq15$ based on
the IR star count model of \cite{nak00}. 
We calculate a total of 280 objects per arcmin$^2$ towards \src, which
suggests a reasonable likelihood of chance positional alignment within an
arcsec. Thus, we cannot place any significant constraints on the nature of
\src\ based on the properties of the proposed counterpart.

\section{Discussion \& conclusions}
\label{disc}

The 2002 and 2006 {\it Chandra}\/ observations represent the only
high-spectral resolution observations of radius-expansion bursts 
from any source to date.
Only six such bursts have been observed in total, and the data from two of
those bursts (from observation \#2748) likely suffer from pileup as well
as having too poor time resolution to resolve the radius-expansion
episode. Nevertheless, this work represents the first such probe into a
poorly-studied burst spectral regime.
We detected no evidence for discrete features in the spectra extracted
from during the radius-expansion episodes, and derive upper limits for
narrow features of approximately $150$~eV.

The super-Eddington flux that drives the radius expansion during
these bursts transforms nearly all the excess luminosity into
kinetic and potential energy of the extended atmosphere. The
radiation-driven winds which result are capable of expelling $\gtrsim 0.5
\%$ of the accreted mass \citep[e.g.][]{hs82,ntl94}. If
a fraction of the heavy elements produced
during burning are expelled in a wind and exposed at the surface,
we may expect to observe 
photoionisation edges from these elements.

However, we found no evidence for photoionization edges, 
and derive typical upper limits of
250--300~eV. The strength of edge features depends upon the elemental
abundances in the burning products, the temperature 
and degree of expansion
of the photosphere
(that sets the degree of ionisation) and the amount of material ejected
\cite[that determines the column depth; e.g.][]{zand10a}. In order to
estimate the likely equivalent widths, we adopted model predictions from
\cite{wbs06}.
We assumed abundances in the burning products 
for the pure-He model (see e.g. Fig. 13 of \citep{wbs06}), since the burst
profiles and other evidence presented here suggest pure He accretion.
The abundance distributions may vary 
if there is
hydrogen present in the ignition layer.
We note that the rp-process burning that would occur if H is present
\cite[e.g.][]{schatz01}
occurs after the radius expansion, so heavy burning products will not be
ejected in the wind, unless they get dredged up from the ashes of previous
bursts.
\cite{zand10a} found evidence for absorption edges with optical depths of
$\tau=0.2$--3 in superexpansion bursts, in which the photospheric radius
increases to $>1000$~km. The expansion in the bursts from 4U~1728$-$34 was
much more modest, at around 20~km (a factor of $\approx2$ above the NS
surface; Fig.
\ref{timespec}), and it is possible that this distinction is the reason
for the nondetection in our data. However, we note that the upper limits
on optical depth that we derive, of $\tau\approx0.2$--0.3, are consistent
with the measured values for the weaker edges in the superexpansion
bursts. Thus, for moderate expansion bursts such as those seen from
4U~1728$-$34, we can rule out the strongest edges detected in the
superexpansion bursts, but not the weakest.

\begin{figure}
 \epsscale{1.15}
 \plotone{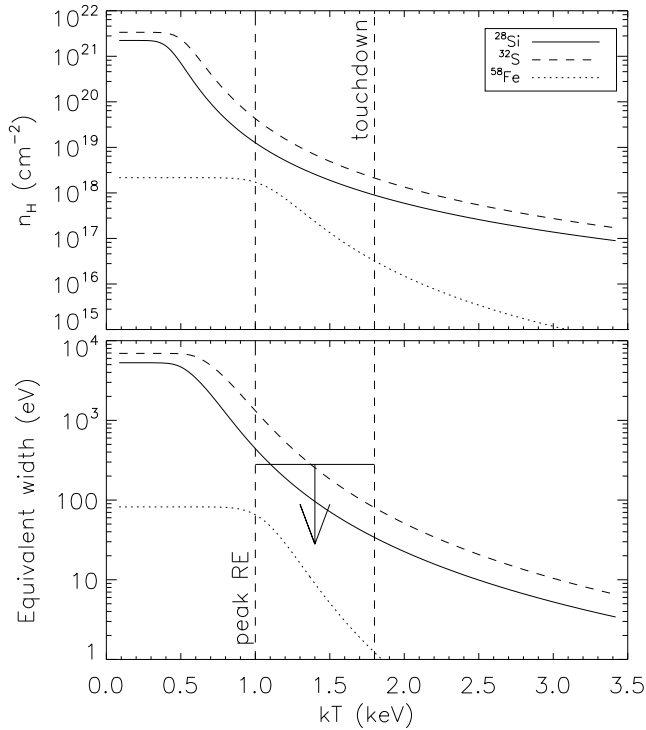}
 \figcaption[]{Column density ({\it top panel}) and edge equivalent width
({\it bottom panel}) for representative elements predicted for model He0.1
(pure He accretion at 0.1~$\dot{M}_{\rm Edd}$) of \cite{wbs06}, as a
function of effective temperature. The estimated effective temperature at
the time of maximum radius expansion and subsequent touchdown during the
first four bursts observed by {\it Chandra}\/ in June 2006 are
indicated ({\it vertical dashed lines}).
The upper limit on edge depth derived for the summed spectra from
four radius-expansion bursts from 4U~1728$-$34 is indicated.
 \label{ew} }
\end{figure}

The predicted column density and equivalent widths for selected elements
are shown in Fig. \ref{ew}. For comparison, we calculated the effective
temperature in the neutron-star photosphere, adopting a correction to the
observed (color) temperature of 2 \cite[see e.g.][]{madej04} and a surface
gravitational redshift of 1.31 (equivalent to a $1.41\,M_\odot$ neutron
star with $R=10$~km), that have the net result of reducing the observed
(color) temperature by a factor of 1.5 to give the effective temperature in
the neutron star frame. We calculated the effective temperatures both at the
time of maximum expansion of the photosphere, and also at the point where
the photosphere is assumed to have rejoined the neutron star surface
(``touchdown''). The detectability of edge features is strongly dependent
upon the effective temperature in the derived range, between 1 and 1.8~keV.
Interestingly, the predicted equivalent width for S and Si at an effective
temperature of $\approx1$~keV are a few hundred to a thousand eV,
comparable to the upper limits on detection derived in the present spectra.
Doppler broadening due to rotation may reduce the detectability
of the edges in moderate radius expansion bursts.
If angular momentum in the expanding photosphere is conserved,
the transverse velocity of the expanded shell in the bursts from
4U~1728$-$34 would be reduced by a factor of two, such that the degree of
Doppler broadening would be equivalent to a source rotating at
$\approx\nu_{\rm spin}/2=181$~Hz.  The reduction in detectability for an
intrinsically broad feature such as an edge will not be so severe as for
spectral lines \cite[e.g.][]{chang06}, although it may be significant.
We hope that future observations, that will contribute additional photons
to the already-accumulated spectrum emitted at the peak of
radius-expansion bursts, 
will allow us to make more sensitive
searches for spectral features.

The properties of the bursts observed from 4U~1728$-$34 by {\it Chandra}\/
suggest that the fuel contains little or no hydrogen.
During the 2006 July observation the source 
entered a state of unusually high
accretion rate, so that (from the second observation, \#6567 onwards) the
burst ignition conditions were reached before sufficient fuel to power a
radius-expansion burst had accumulated. The occurrence of bursts with
characteristic He-rich profiles and high $\alpha$-values at 
such short recurrence times are
difficult to understand based on standard burst theory.
For a neutron star accreting mixed H/He at
around 10\% the Eddington rate (within the range in
which we observed \src), H will burn stably between the bursts, reducing
the fraction in the fuel at ignition. However, 
the burst recurrence time observed in 2006 July is insufficient to exhaust
the H for solar abundances. Steady (hot CNO) H-burning is mediated by
$\beta$-decays which limit the rate, so that the
accreted hydrogen will only be exhausted at the ignition depth in a time
$t=11(X_0/0.7)(Z_{\rm CNO}/0.02)^{-1}$~hr, where $X_0$ is the hydrogen fraction in the
accreted material and $Z_{\rm CNO}$ is the mass fraction of CNO nuclei (0.7 and
0.02 respectively, for solar composition). For burst recurrence times of
$\approx2$~hr there is insufficient time between bursts to reduce the H-fraction
in the fuel significantly below the accreted level prior to burst
ignition. If the source were accreting hydrogen at approximately solar
composition, the expected burst profiles would be much longer, with
$\approx5$~s
rises and tails out to 100~s, similar to what has been observed for
GS~1826$-$24 \cite[]{gal03d}. In those bursts the longer timescale comes
from the much slower progression of $\beta$-decay mediated rp-process H-burning. 
We suggest three  possible explanations for the burst properties. First,
that the source accretes material which is unusually rich in CNO nuclei.
With $Z_{\rm CNO}$ approximately 5 times in excess of the solar value, the time to
exhaust all the H would be of order the burst recurrence time observed
during the 2006 {\it Chandra}\/ observation. There is no observational support for
this suggestion, as the optical counterpart for the source has not been
confirmed, so the type of the mass donor is not known. Second, it is
possible that shear-mediated mixing allows CNO nuclei from the ashes of
previous bursts to be dredged up into the freshly accreted material, as
has been predicted for slowly rotating neutron stars by \cite{pb07}.
\src\ is known to be rotating at 363~Hz \cite[]{stroh96}, which is fairly slow for
accreting neutron-stars
in LMXBs, but the full effects of such mixing in
sources accreting hydrogen as well as helium have not yet been fully
investigated. Either of these possibilities leave open the chance of long
H-rich bursts occurring occasionally from the source, although these have
not yet been observed. Continued
observations, as well as an archival search underway with data from
multiple instruments, will test these hypotheses.

Third, it remains possible that the source accretes H-poor material, from
an evolved (e.g. white dwarf) donor. Such systems are referred to as
ultracompact binaries, due to their typically very short ($<80$~min) orbital
periods 
\cite[e.g.][]{zjm07}.
The 10.77~min modulation in the intensity of low-energy X-rays detected
for the first time by Chandra from \src\ 
supports this
possibility. If this modulation was orbital, the system would be
essentially a twin of 
4U~1820$-$30.
Like that source, \src\ will be detectable with the ESA/NASA Laser
Interferometer Space Antenna \cite[LISA;][]{nelemans08}.
Existing measurements of
the candidate IR counterpart 
cannot significantly constrain this
possibility, 
and deeper infrared observations are required to unambiguously confirm the
ultracompact nature of this source,
via detection of a corresponding modulation in the
counterpart intensity.

\acknowledgments

This research has made use of data obtained from the Chandra Data Archive
and the Chandra Source Catalog, and software provided by the Chandra X-ray
Center (CXC) in the application package CIAO. 
We thank D. Kaplan for
supplying the code to estimate star counts in the 4U~1728$-$34 field.

Facilities: \facility{CXO(ASIS)}, \facility{XMM(EPIC)}


\clearpage

\clearpage

\end{document}